\documentclass[12pt]{article}
\usepackage{graphicx}
\usepackage{float}
\usepackage{graphicx}
\begin{document}
\rightline{NKU-2017-SF1}
\bigskip

%%%%%%%%%%%%%newcommands%%%%%%%%%%%

\newcommand{\be}{\begin{equation}}
\newcommand{\ee}{\end{equation}}
\newcommand{\noi}{\noindent}
\newcommand{\refb}[1]{(\ref{#1})}
\newcommand{\ra}{\rightarrow}
\newcommand{\bi}{\bibitem}
\newcommand{\bff}{\begin{figure}}
\newcommand{\eff}{\end{figure}}

%%%%%%%%%%%%%title%%%%%%%%%%%%%%%%%%%

\begin{center}
{\Large\bf Spinning dilaton black hole in 2+1 dimensions as a particle accelerator}
\end{center}

\hspace{0.4cm}

\begin{center}
Sharmanthie Fernando \footnote{fernando@nku.edu}\\
{\small\it Department of Physics, Geology \& Engineering Technology}\\
{\small\it Northern Kentucky University\\
Highland Heights\\
Kentucky 41099\\
U.S.A.}\\

\end{center}

%%%%%%%%%%%%%%%%Abstract%%%%%%%%%%%%%%%%%%%%%
\begin{center}
{\bf Abstract}
\end{center}

In this paper we have studied  particle collision around a  spinning dilaton black hole in 2 +1 dimensions. This black hole is a solution to the low energy string theory in 2+1 dimensions. Time-like geodesics are presented in detail and the center of mass energy of two particles collision at the horizon of a spinning black hole is considered. We noticed that there is a possibility of the two masses to create infinite center of mass energy. 

\hspace{0.5cm} 

{\it Key words}: spinning, accelerator, black hole, dilaton, strings

%%%%%%%%%%%%%%%%%%%%%%%%%%%%%%%%%%%

\section{Introduction}

%%%%%%%%%%%%black holes as accelerators%%%%%%%%%%%%

Ban$\tilde{a}$dos, Silk and West \cite{ban} demonstrated that two particles colliding near the horizons of a Kerr black could have large center of mass (CM) energy: for this process to occur, the black hole has to be extremal and one of the particles has to have critical angular momentum. Now, this process, well known as BSW effect has its draw back as pointed out by two papers, one by Berti et.al. \cite{ema} and the other by Jacobson and Sotirion \cite{ted}: there are astrophysical limitations for the Kerr black hole to become extreme and gravitational radiation and the back reaction must be taken into account in the process. Many black holes, both rotating and charged, have been studied in this context.

Grib and Pavlov showed that scattering energy of the particles in the vicinity of a rotating black hole can reach very large values not only for extreme rotating black holes but also for non extreme rotating black holes \cite{grib}. Kerr-de Sitter black hole could act as a particle accelerator and create unlimited CM energy for two particle collisions even if it is not extremal \cite{li}. In the BSW effect, one of the particles should have critical angular momentum; Harada and Kimura discussed that if the two particles collide at the inner-most-stable-circular-orbit (ISCO), then this fine tuning will occur naturally \cite{har}. Particle  acceleration in the back ground of  the Kerr-Taub-NUT space-time has been studied by Liu et.al \cite{liu}: there, the CM energy can reach infinite for extremal black holes. Collisions of two particles in the background of a rotating black hole in string theory, well known as the Sen black hole were studied by Wei et.al. \cite{wei4}.

Extension of the BSW effect to charged black holes with charged particles were done in \cite{zhou2}. The BSW mechanism was shown to exist in rotating cylindrical black holes \cite{said}.   Not only rotating black holes, but non rotating charged black holes could act as particle accelerators when they are extreme as pointed out by  Zaslavskill   \cite{oleg}. This phenomena where extreme charged black holes with charged particles can act as particle accelerators is true for string black holes  \cite{fernando1} and in Einstein-Maxwell-dilaton black holes  \cite{mao}. Extending this idea, charged black holes as particle accelerators were studied by Wei et.al \cite{wei7}. In another interesting work, particle accelerators inside spinning black hole was considered by Lake \cite{lake}. Not only spinless particles, collision of spinning particles have been considered recently \cite{oleg2} \cite{guo} \cite{zhang}.

Low dimensional gravity provides a simpler setting to investigate properties of black holes such as particle acceleration. For example, the famous BTZ black hole in 2 + 1 dimensions \cite{banados} \cite{ban2} has been studied in variety of context to understand properties of black holes which are otherwise mathematically challenging. Our goal in this paper is to study particle acceleration around a spinning dilaton black hole in 2+1 dimensions. The first dilaton black hole in 2 + 1 dimensions which was static and charged was derived by Chan and Mann in\cite{chan1}. As an extensions to that work  a class of spinning dilaton black holes in 2 +1 dimensions with a dilaton  were derived by  Chan and Mann in \cite{chan2}.  For certain parameters of the theory,  one of such black holes corresponds to the  low-energy string action. Our focus in this paper is the one corresponding to the string action. There are several other noteworthy work related to dilaton black holes in 2+1 dimensions: Modifications of the BTZ black hole by a dilaton field was presented in  \cite{chan3}. Chen generated new class of  dilaton solutions by applying T-duality to existing  solutions in 2+1 dimensions \cite{chen}. By compactification   of 4D cylindrical solutions, rotating dilaton solutions were generated  by Fernando \cite{fer1}.  There are very few papers on the particle acceleration in low dimensions. The famous rotating BTZ black hole \cite{banados}  was studied in this context by  Yang et.al.\cite{yang}. Particle acceleration of charged hairy black hole in 2+1 dimensions were studied by Sadeghi et.al. \cite{sad}. 
 
The paper is organized as follows: in section 2, we will present the details of the spinning dilaton black hole in 2+1 dimensions. In section 3, time-like geodesics are presented. Analysis of the effective potential is done in section 4 and the center of mass is discussed in section 5. Finally the conclusion is given in section 6.

%%%%%%%%%%%%%%%%%%%%%%%%%

\section{ Introduction to the spinning dilaton  black hole in 2+1 dimensions}

In this section we will present the geometry and other properties of the spinning dilaton black hole in 2+1 dimensions.  Chan and Mann \cite{chan2} derived neutral spinning black hole solutions by considering the following action:
\begin{equation}
S = \int d^3x \sqrt{-g} \left[ R - 4  (\bigtriangledown \phi )^2  + 2 e^{b \phi} \Lambda \right]
\end{equation}
Here, $ \Lambda$ is  the cosmological constant, $\phi$ is the dilaton field, and $R$ is the Ricci scalar.  In  this paper  $ \Lambda > 0 $ corresponds to anti-de-Sitter space and  $\Lambda < 0$ corresponds to the de-Sitter space.  The metric of the corresponding solution is,
$$ds^2 = - \left( \frac{ 8 \Lambda r^N}{ (3 N -2)N} + p r^{ 1 - \frac{N}{2}}\right) dt^2 + \frac{dr^2}{\left[ \frac{ 8 \Lambda r^N}{ (3 N -2)N} + \left( p - \frac{ 2 \Lambda \alpha^2}{ ( 3 N -2)N p} \right) r^{ 1 - \frac{N}{2}}\right]}$$
\begin{equation} \label{metric1}
- \alpha r^{ 1 - \frac{N}{2}} dt d\theta + \left( r^N - \frac{\alpha^2}{ 4 p} r^{ 1 - \frac{N}{2}} \right) d \theta^2
\end{equation}
The mass $M$ and the angular momentum J of the solution is given by,
\begin{equation}
M = \frac{N}{2} \left[\frac{2 \Lambda \alpha^2}{ ( 3 N -2)N p} \left( \frac{4}{N} - 3 \right) - p \right]; \hspace{0.2cm} J = \frac{ 3 N - 2}{ 4} \alpha
\end{equation}
$p$ and $\alpha$ are integration constants and are given by,
\begin{equation}
p = -\frac { M} { N} - \sqrt{\frac{M^2}{N^2} + \left( \frac{4}{N} - 3 \right) \frac{ 2 \Lambda \alpha^2}{ (3 N -2)N} }
\end{equation}
To avoid closed-time-like co-ordinates, the integration constant $p < 0 $ must be chosen to be negative. The dilaton field $\phi$ is given by,
\begin{equation}
\phi = k ln( r)
\end{equation}
Here $k$ and $b$ are  related to $N$ as,
\begin{equation}
k = \pm \frac{1}{4} \sqrt{ N ( 2 - N)} ; \hspace{1 cm} bk = N -2
\end{equation}

In the paper by Chan and Mann  \cite{chan2} it was stated that positive mass ($M > 0$) black holes exists only for  $\Lambda > 0$ and $ \frac{2}{3} \leq N \leq 2$. 
In this paper, we will focus on a special class of black holes with $ b = 4$, $ N =1$ and $ k = -1/4$. Such values lead to the low-energy string effective action,
\begin{equation} \label{action2}
S = \int d^3x \sqrt{-g} \left[ R - 4  (\bigtriangledown \phi )^2  + 2 e^{ 4 \phi} \Lambda \right]
\end{equation}
The action in eq.$\refb{action2}$ is related to the low-energy string action in 2+1 dimensions by a conformal transformation given as follows,
\begin{equation}
g_{\mu \nu}^{String} = e^{ 4 \phi} g_{\mu \nu}^{Einstein} 
\end{equation}
Now, the spinning black hole solution corresponding to the  action in eq.$\refb{action2}$  is given by,
\begin{equation} \label{metric3}
ds^2= - f(r)  dt^2 +  \frac{dr^2}{ h(r) } - 4 J r d \theta dt + R(r)^2 d \theta^2
\end{equation}
where,
\begin{equation}
f(r) =  \left(8 \Lambda r^2  - ( M + \sqrt{ M^2 + 32 \Lambda J^2})r\right)
\end{equation}
\begin{equation}
h(r) = \frac{(8 \Lambda r^2  - 2 Mr )}{ 4 r^2 }
\end{equation}
\begin{equation}
R(r)^2 = \left(   r^2 + \frac{( - M + \sqrt{ M^2 + 32 \Lambda J^2} ) }{8 \Lambda} r \right)
\end{equation}
Later in the paper, we will take 
\be
R(r)^2 = r^2 + a r
\ee
 with 
\be \label{avalue}
a = \frac{( - M + \sqrt{ M^2 + 32 \Lambda J^2} ) }{8 \Lambda}
\ee
 to facilitate computations. The dilaton field is given by,
\begin{equation}
\phi = -\frac{1}{4} ln(r)
\end{equation}
In deriving the  metric in eq$\refb{metric3}$  from the one in eq$\refb{metric1}$, the value of $\alpha = 4 J$ is substituted. Also, a simple coordinate transformation $ r \rightarrow r^2 $ is done. With the chosen constants,  $p$
is computed as,
\begin{equation}
p = - M - \sqrt{ M^2 + 32 \Lambda J^2}
\end{equation}
The location of the event horizon is given by $h(r)=0$ at,
\begin{equation}
r_h =  \frac{M}{4 \Lambda} 
\end{equation}
and there is  a singularity at $r=0$. The Ricci scalar and Kretschmann scalars diverge only at $r =0$. 

The Hawking temperature of the black hole is given by,
\begin{equation}
T_H= \frac{1}{4 \pi} \left( \frac{1}{16 \Lambda^2}  + \frac{J^2}{ M \Lambda( M + \sqrt{ M^2 + 32 \Lambda J^2}) }  \right)^{-1/2}
\end{equation}
Quasinormal modes and area spectrum  of scalar perturbations of the above black hole  were studied by the current author \cite{fernando5}.
 
%%%%%%%%%%%%%%%%%%%%%%%%%%%%%%%%%%%%%%%%

\section{ Time-like geodesics of the test particles of the dilation black hole}

%%%%%%%%%%%%%%%%%%%%%%%%%%%%%%%%%%%%%%%

In this section we will present  time-like geodesics of the spinning dilaton black hole. To derive the geodesics, the formalism in Chandrasekhar's book is followed \cite{chandra}.  The Lagrangian of the massive test particle in this black hole background is given by,
\be \label{lag}
{\cal{L}} =  -\frac{1}{2} \left( g_{\mu \nu} \frac{ dx^{\mu}}{d \tau} \frac{ dx^{\nu}}{ d \tau} \right) 
\ee
\be \label{lag2}
= - \frac{1}{2} \left( - f(r) \left( \frac{dt}{d\tau} \right)^2 +  \frac{1}{h(r)}\left( \frac{dr}{d \tau} \right)^2 + R(r)^2 \left(\frac{d \phi}{d \tau} \right)^2  - 4 J r  \left(\frac{ dt}{ d \tau}\right) \left(\frac{d \phi }{d \tau} \right) \right)
\ee
Here,  the parameter $\tau$ is the proper time for massive particles.  The metric functions $g_{\mu \nu}$ in eq$\refb{lag}$ corresponds to,
\be
g_{tt} = -f(r); \hspace{1 cm} g_{rr} = \frac{1}{ h(r)}; \hspace{1 cm} g_{\phi \phi} = R^2(r); \hspace{1 cm} g_{t \phi} = - 2 J r
\ee
Each coordinate in the space-time has a corresponding canonical momenta $p_{\mu}$ given by,
\begin{equation} \label{pt}
p_t =  - g_{tt} \dot{t} - g_{t \phi} \dot{\phi}
\end{equation}
\begin{equation} \label{phi}
p_{\phi}  =    g_{\phi \phi} \dot{\phi} + g_{t \phi} \dot{t}
\end{equation}
\begin{equation}
p_{r} = g_{rr} \dot{r}
\end{equation}
The spinning dilaton  black hole  have two Killing vectors $\partial_t$ and $ \partial_{\phi}$. Hence, the canonical momenta  $p_t$ and $p_{\phi}$ are conserved: these constants  of the particle  are labeled as energy per unit mass $E$ and angular moment per unit mass $L$.   From eq.$\refb{pt}$ and eq.$\refb{phi}$,  $E$ and $L$  can be solved to be,
 \begin{equation} \label{pt2}
E = f(r) \dot{t} +  2 J r  \dot{\phi} 
\end{equation}
\begin{equation} \label{pphi2}
L=   R^2(r)  \dot{\phi} - 2 Jr  \dot{t}
\end{equation}
Eq$\refb{pt2}$ and eq$\refb{pphi2}$ are solved to obtain $\dot{t}$ and $\dot{\phi}$ as,
\be \label{pt3}
\dot{t} = \frac{ \left( -2  L J r+  E R^2(r) \right)} { \left(4 J^2 r^2 +  f(r) R^2(r)\right)}
\ee
\be
\dot{\phi} = \frac{  \left(2 E J r +  f(r) L\right) }{ \left(4 J^2 r^2 +  f(r) R^2(r)\right)}
\ee
Using the identity,
\be
f(r) = \frac{ (4 r^4 h(r) - 4 r^2J^2)}{ R^2(r)}
\ee
$\dot{t}$ and $\dot{\phi}$ can be rewritten as,
\be \label{tdot4}
\dot{t} = \frac{ \left( -2 L J r +  E R^2(r) \right)} { 4 r^4 h(r)} = u^t
\ee
\be \label{pdot4}
\dot{\phi} = \frac{  \left( 2 E J r +  f(r) L\right) }{  4 r^4 h(r)} = u^{\phi}
\ee
We will assume $\dot{t} > 0$ for all $r > r_h$ so that the motion is forward in time outside the horizon. Hence,
\be \label{con}
 \left(E R^2(r) - 2 rL J \right)  > 0,  \hspace{1 cm} \forall \hspace{0.1 cm}  r > r_h
\ee
The four velocity of the particles are given by $u^{\mu} = \frac{ dx^{\mu}}{d \tau}$ and they are normalized as, $u^{\mu} u_{\mu} = -1$. $u^{t} = \dot{t}$ and $ u^{\phi}= \dot{\phi}$ are already derived above. The normalized condtion is expresses as
\be \label{nor}
g_{tt} (u^t)^2 + g_{rr}(u^r)^2 + g_{\phi \phi} (u^{\phi})^2 + 2 g_{\phi t} (u^{\phi}) ( u^t) = -1
\ee
By substituting $u^t$ and $u^{\phi}$ from eq$\refb{tdot4}$ and eq$\refb{pdot4}$, one can obtain   $u^r= \dot{r}$ as,
\be \label{rdot}
 \dot{r}^2   =  \frac{ - \left( 1 + g_{tt} \dot{t}^2 + g_{\phi \phi} \dot{\phi}^2 + 2 g_{t \phi} \dot{t} \dot{ \phi} \right) }{ g_{rr} } 
\ee
which is simplified to,
\be \label{rdot}
\dot{r}^2 = \frac{1}{ 4 R^2(r) r^4} \left( - 4 r^4 h(r) ( R^2(r) + L^2) + ( 2 r L J -  E R^2(r))^2\right)
\ee
By identifying,
\be
K = ( - 2 r J L +  E R^2(r))
\ee
and
\be
H^2 = K^2 - 4 r^4 h(r) ( R^2(r) + L^2)
\ee
$\dot{r}$ can be written in a short form as,
\be
\dot{r} = -  \frac{ H}{ 2 R(r) r^2} = u^r
\ee
The minus sign is chosen since we assume the particle is  falling towards the black hole.

%%%%%%%%%%%%%%%%%%%%%%%%%%

\section{ Analysis of the effective potential for the spinning dilaton black hole}

\begin{figure} [H]
\begin{center}
\includegraphics{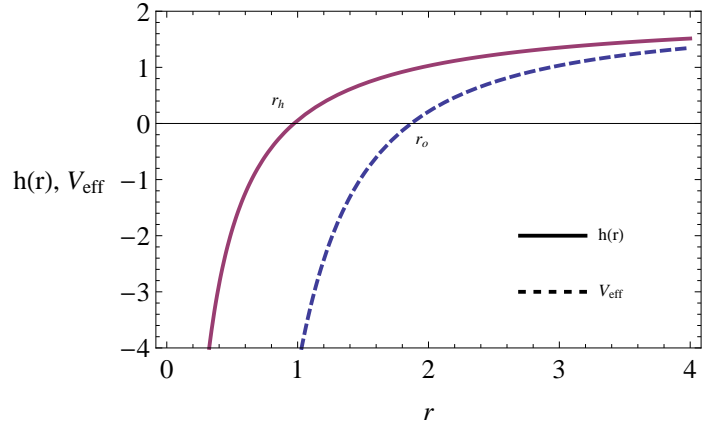}
\caption{The figure shows  $h(r)$ and $V_{eff}$ vs $r$. Here $ M = 3.89, J = 2.94, E = 2.92, \Lambda =1$ and $ L = 0.15$}
\label{pot1}
\end{center}
\end{figure}

\begin{figure} [H]
\begin{center}
\includegraphics{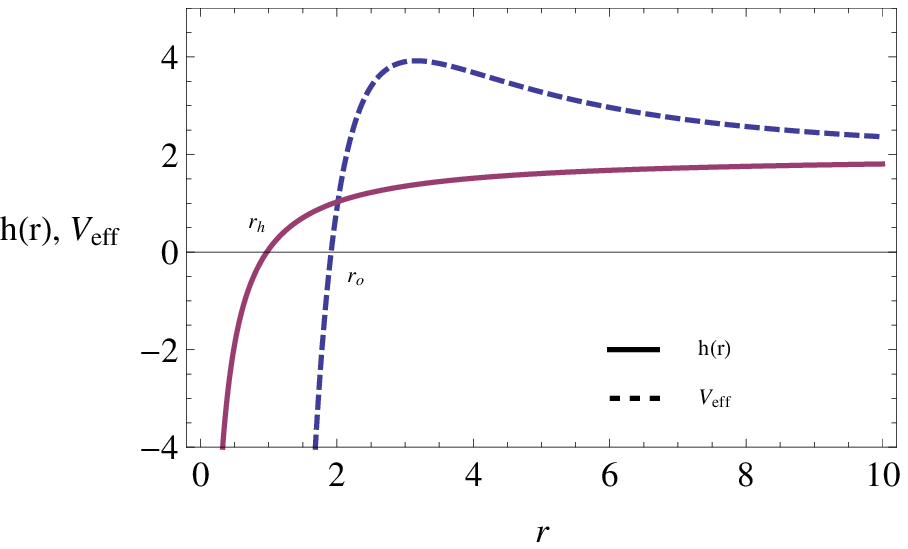}
\caption{The figure shows  $h(r)$ and $V_{eff}$ vs $r$. Here $ M = 3.89, J = 2.94, E = 2.92, \Lambda =1$ and $ L = 6 $}
\label{pot2}
\end{center}
\end{figure}

\begin{figure} [H]
\begin{center}
\includegraphics{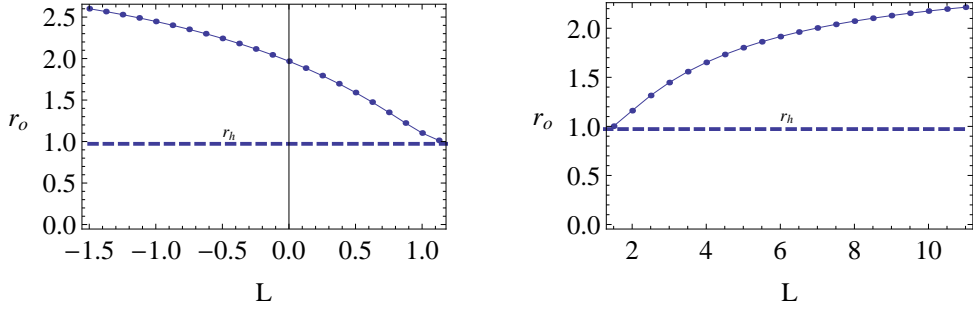}
\caption{The figure shows  $r_o$ vs $L$. Here $ M = 3.89, J = 2.94, E = 2.92, \Lambda =1$}
\label{roots}
\end{center}
\end{figure}

In this section the effective potential is analyzed to see if a particle could reach the horizon. The effective potential is calculated as,
\be 
\dot{r}^2  + V_{eff} =0
\ee
Hence by substituting from eq$\refb{rdot}$, one obtain $V_{eff}$ as,
\be
V_{eff} =  \frac{  4 r^4 h(r) ( R^2(r) + L^2) - ( 2 r L J -  E R^2(r))^2}{ 4 R^2(r) r^4}
\ee
Since $\dot{r} = \sqrt{- V_{eff} }$, the particle can exists only in the regions where $V_{eff} < 0$. For large $r$, $V_{eff} \ra 2 \Lambda$. In Fig$\refb{pot1}$ and Fig$\refb{pot2}$, $V_{eff}$ and $h(r)$ are plotted. Both figures have the same parameters except $L$.  Fig$\refb{pot1}$ has a smaller $L$.  As one can see $V_{eff}$ has only one root  for $ r > 0$ in both cases. Only difference is that for larger $L$, $V_{eff}$ has a maximum outside the root. The root of $V_{eff}$, given by $r_o$,  lies outside the horizon $r_h$. One can prove that this is the case for all parameters as follows: At $ r = r_h$, $V_{eff}$ is given by,
\be \label{poth}
V_{eff} =  \frac{  - ( 2 r_h L J -  E R^2(r_h))^2}{ 4 R^2(r_h) r_h^4} = - \frac{ K(r_h)^2} { 4 R^2(r_h) r_h^4 }
\ee
At $r= r_h$, $R(r_h)$ is given by,
\be
R(r_h) = \frac{ M ( M + \sqrt{ M^2 + 32 J^2 \Lambda})}{ 32 \Lambda^2}
\ee
Hence $R(r_h)^2 > 0$ at $r = r_h$. In fact  $R(r)$ greater than zero for all $ r > 0$. Hence $V_{eff} \leq 0$ at the horizon. $V_{eff}(r_h) =0$ for a critical angular momentum given by,
\be
L_c = \frac{ E}{ 16 \Lambda J} ( M + \sqrt{ M^2 + 32 J^2 \Lambda} )
\ee
Notice that when $V_{eff}(r_h) = 0$ implies $K(r_h) =0$. We  computed the root of $V_{eff}$ for range of angular momentum and plotted in Fig$\refb{roots}$. One can see that the root $r_o \geq 0$ for all $|L| > L_c$. 
Hence if $|L| > L_c$, the particle can start at rest from a finite distance away from the horizon and fall towards the horizon and eventually fall inside the black hole. The larger the $L$, further away can the particle start its motion.
 
%%%%%%%%%%%%%%%%%%%%%

\section{ Center of mass energy for two particle collision}

Now that we have established the fact that the particles can fall towards the black hole event horizon for a given set of parameters,  we would like to compute the center of mass energy for two massive particles in the black hole background. Let  the  four velocities of the particles be  $u_1^{\mu}$ and $u_2^{\mu}$. We will assume both have the same rest mass $m_0$. Then, the  center of mass  energy is given by,
\be
\hat{E}_{CM} = 2 m_0^2 ( 1 - g_{\mu \nu} u_1^{\mu} u_2^{\nu} )
\ee
In the rest of the paper we will calculate $E_{cm} = \hat{E}_{cm}/2$ with $m_0=1$, which is given by,
\be
E_{CM} = \frac{ \left(4 r^4 h(r) ( R^2(r) - L_1 L_2) + ( K_1 K_2 - H_1 H_2) \right)} { 4 R^2(r) r^4 h(r)}
\ee
where,
\be
K_1 = - 2 r J L_1 + E_1 R^2(r)
\ee
\be
K_2 = - 2 r J L_2 + E_2 R^2(r)
\ee
When the particles reach  the horizon at $ r = r_{h}$, $h(r) \ra 0$, $ H_1 \ra \sqrt{K_1^2}$, and $ H_2 \ra \sqrt{K_2^2}$. Therefore,
\be \label{energy}
E_{CM}(r \ra  r_h ) = \frac{ 1}{ 4 R^2(r) r^4 h(r) } \left( K_1 K_2 - \sqrt{K_1^2}  \sqrt{K_2^2} \right)
\ee
At the horizon, $h(r) =0$, hence the denominator of eq.$\refb{energy}$ is zero. Due to the condition given in eq.$\refb{con}$, $K_1, K_2  \geq 0$ at the horizon. Hence the numerator is also zero leading to an undetermined value for $E_{CM}$. However, one can use the L' Hospital's  rule to calculate the limiting value of eq$\refb{energy}$ as $ r \ra r_h$. The result is given by,
\be \label{limit}
E_{CM} = \frac{ a^2 r_h^2 A_1 + a r_h A_2 + A_3}{ 2 K_1(r_h) K_2(r_h)}
\ee
where
\be
A_1 = (E_1 + E_2)^2 
\ee
\be
A_2 = 2(E_1 + E_2)^2 r_h^2 + ( E_2 L_1 - E_1L_2)^2 - 4 r_h J (E_1 + E_2) ( L_1 + L_2) 
\ee
\be
A_3 = r_h^2 \left( 4 J^2 ( L_1 + L_2)^2 - 4 J r_h ( L_1 + L_2) ( E_1 + E_2) +   ( E_1 + E_2)^2 r_h^2 + ( E_1 L_2 - E_2 L_1)^2 \right)
\ee
Here $a$ is the value given in eq$\refb{avalue}$.

\begin{figure} [H]
\begin{center}
\includegraphics{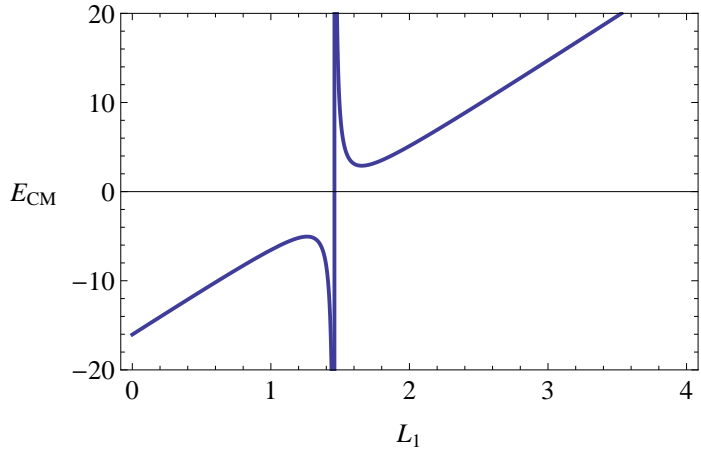}
\caption{The figure shows  $E_{CM}$ vs $L_1$. Here $ M = 1, J = 1, \Lambda =1, E_1 = 3.46, E_2 = 0.71, L_2 = 0.37$}
\label{ecm}
\end{center}
\end{figure}

The numerator of eq$\refb{limit}$ is finite  at $ r = r_h$. So, it can be seen that when $K_1(r_h) =0$ or $K_2(r_h) =0$, the center of mass energy becomes infinity. When one solve $K_1(r_h) =0$ or $K_2(r_h) =0$,  one obtain the critical angular momentum $L_{c1}$ or $L_{c2}$ which was discussed in section(4).  $E_{CM}$ is plotted for varying $L_1$ with all other parameters fixed as in Fig$\refb{ecm}$. In order to get infinite center of mass energy, particle number 1 reach the critical angular momentum as shown in the figure. Notice that the limiting value of the critical angular momentum has to be reached from the right of the asymptotic value on order to get a positive center of mass energy.

We want to mention that if both $K_1(r_h)$ and $K_2(r_h)$ are zero, then $E_{CM}$ is finite.  One can prove it as follows: When $K_1(r_h) = K_2(r_h) =0$, it implies that  $H_1(r_h) = H_2(r_h) =0$. Hence for this special case,
\be
E_{CM} = \frac{ \left(4 r^4 h(r) ( R^2(r) - L_1 L_2)  \right)} { 4 R^2(r) r^4 h(r)} = 1 - \frac{ L_1 L_2} { R^2(r_h)}
\ee
which is clearly finite. Therefore, in order to obtain infinite center of mass energy, only one of the colliding particles should have the critical angular momentum.

%%%%%%%%%%%%%%%%%%%%%%%%%%%

\section{Conclusions}

In this paper we have studied a spinning dilaton black hole in 2+1 dimensions. This black hole is a solution to low-energy string action in 2+1 dimensions. It has a single horizon at $r_h = \frac{M}{ 4 \Lambda}$: this is in contrast to many spinning black hole  which has two horizons.

We have studied the time-like geodesics in detail. We have calculated the three velocities $u^{\mu}$. The effective potential reaches a constant value $2 \Lambda$ for large $r$ and has a zero at $r= r_o$. Th effective potential  is negative for $r < r_o$. Hence, a particle cannot exist classically for $r > r_0$. The value of $r_o$ is larger for large $L$. Also, $ r_o > r_h$:  a particle starting at rest from a finite value $r < r_o$ will fall into the black hole.

Our main goal in this paper has to  study the two particle collisions and to see if the center of mass could reach very high values. In fact we noticed that it is possible to generate infinite center of mass energy if the particles collide closer to the horizon $r_h$. Therefore, the spinning dilaton black hole could act as a particle accelerator. In extending this work it would be interesting to study the charged dilaton black hole in 2+ 1 dimensions \cite{chan1}  with charged particles to see if they could generate high center of mass energy.

%%%%%%%%%%%%%%%%%%%%%%%%%

%%%%%%%%%%%%%%

\end{document}